# Determination of Gd concentration profile in $UO_2$-$Gd_2O_3$ fuel pellets

D. Tobia [a,1], E. L. Winkler [a], J. Milano [a], A. Butera [a], R. Kempf [b], L. Bianchi [c], F. Kaufmann [c]

[a] *Laboratorio de Resonancias Magnéticas, Centro Atómico Bariloche – CNEA and CONICET, (8400) S.C. de Bariloche, Argentina*
[b] *División Caracterización de Combustibles Avanzados, Gerencia Ciclo Combustible Nuclear, Centro Atómico Constituyentes – CNEA, (1650) San Martín, Pcia. de Buenos Aires, Argentina*
[c] *Departamento de Combustibles Avanzados, Gerencia Ciclo Combustible Nuclear, Centro Atómico Constituyentes – CNEA, (1650) San Martín, Pcia. de Buenos Aires, Argentina*

## Abstract

A transversal mapping of the Gd concentration was measured in $UO_2$-$Gd_2O_3$ nuclear fuel pellets by electron paramagnetic resonance spectroscopy (EPR). The quantification was made from the comparison with a $Gd_2O_3$ reference sample. The nominal concentration in the pellets is $UO_2$: 7.5 % $Gd_2O_3$. A concentration gradient was found, which indicates that the $Gd_2O_3$ amount diminishes towards the edges of the pellets. The concentration varies from (9.3±0.5)% in the center to (5.8±0.3)% in one of the edges. The method was found to be particularly suitable for the precise mapping of the distribution of $Gd^{3+}$ ions in the $UO_2$ matrix.

## 1-Introduction

One of the major challenges of the nuclear industry is to improve the performance, safety and lifetime of reactors [1]. In this area there are many efforts focused on the research and development of new materials in order to extend the fuel lifetime, increase the burn-up and optimize the power density distribution. With this aim a neutron absorber material is usually incorporated into the $UO_2$ nuclear fuel. Gadolinium is an excellent burnable poison because it presents a large cross section for neutron absorption and allows the compensation of the excess reactivity of the fuel in the beginning of its life. The solid solution $(U,Gd)O_2$ can be fabricated by several processes and the material presents different physical and chemical properties depending on the Gd concentration and the synthesis route. The study of the synthesis densification process [2,3] and phase homogeneity [4-8] is crucial because the presence of micropores or an inhomogeneous distribution of Gd ions could cause internal cracks and/or affect the fuel performance. Furthermore, it is essential to know the Gd distribution profile in the fuel pellets since this parameter modifies the fuel reactivity and, if it goes out of the calculated range, the reactor design could not be fulfilled.

---

[1] Corresponding author: Dina Tobia
E-mail: dina.tobia@cab.cnea.gov.ar
Address: Av. Bustillo 9500, (8400) S.C. de Bariloche, RN, Argentina
Phone: ++54 +294 4445158

Within this frame, it is crucial the search of new and more precise analysis techniques that allow a better characterization of Gd content and distribution in these fuel pellets [9,10].

Recently, a photothermal photodeflection technique was successfully employed to map the homogeneity in ceramic samples of nuclear interest. [9] The two-dimensional mapping of the thermal diffusivity allowed to quantify and measure the microscopic and mesoscopic pores in the samples. Also, clusters of urania grains with low or null gadolinium content were identified in this manner. Although this technique provides excellent quantitative measures for the homogeneity of these samples at microscopic levels, a standardized protocol involving a direct determination of Gd concentration along the $(U,Gd)O_2$ pellets is still lacking.

Electron paramagnetic resonance spectroscopy (EPR) emerges as an excellent candidate to carry out this goal. Due to its high sensitivity (~$10^{10}$ spins/Gauss.Hz$^{1/2}$ [11]), one of the applications of EPR is the quantification of the number of paramagnetic centers of a material. [12, 13] The quantification is performed taking into account that the EPR absorption intensity is proportional to the number of magnetic ions of the system [14]. For example, the EPR spectroscopy is a standard technique to measure the irradiation dose (in the 1-$10^5$ Gy range [15]) by the quantification of paramagnetic centers generated by irradiation in crystalline alanine [16, 17, 18]. In the case of $UO_2$-$Gd_2O_3$, one of the major advantages is that the $Gd^{3+}$ EPR signal is particularly intense due to its electronic configuration (orbital angular moment $L$=0, spin $S$=7/2) [19]. This fact allows to quantify the concentration of Gd ions through a direct comparison of the EPR intensity of the $UO_2$-$Gd_2O_3$ sample with a standard sample with a known concentration of magnetic ions. In this work we report a method to measure with a very high precision the Gd concentration profile in a $UO_2$-$Gd_2O_3$ pellet by EPR spectroscopy.

## 2. Experimental
## 2.1. Electron Paramagnetic Resonance Technique

The electron paramagnetic resonance corresponds to the resonant absorption of microwave radiation by paramagnetic ions in a static magnetic field. From the EPR absorption three parameters are usually determined: the resonance field ($H_r$), the linewidth ($\Delta H$), and the intensity ($I_{EPR}$). The EPR absorption is centered at the resonance field $H_r$ which is related to the gyromagnetic factor, $g$, that depends essentially on the electronic configuration of the ion. The linewidth is related to the spin relaxation mechanism. Finally, the spectrum intensity could be determined from the area under the EPR absorption curve and is proportional to the static magnetic susceptibility, $\chi_{DC}$, when all the magnetic ions contribute to the resonance [13, 20]:

$$I_{EPR} \propto V\chi_{DC} = \frac{Ng^2\mu_B^2 S(S+1)}{3k_B T} \quad (1)$$

where $V$ is the sample volume, $N$ is the total number of magnetic ions, $g$ is the gyromagnetic ratio, $\mu_B$ is the Bohr magneton, $k_B$ is the Boltzmann constant and $T$ is the temperature. Because the EPR spectrometer records the derivative of the absorption line as the magnetic field is swept, $I_{EPR}$ is obtained from the second integration of the measured spectrum. The ratio of the integrated intensity $I_{EPR}$ of two samples is ideally independent of the spectrometer condition. Therefore, from a comparison with a reference sample the concentration of magnetic ions could be determined:

$$I_{EPR\_A} / I_{EPR\_O} = \frac{N_A g_A^2 S_A(S_A+1)}{N_O g_O^2 S_O(S_O+1)} \quad (2)$$

where the test sample with an unknown number of magnetic ions is named $A$ and the label $O$ corresponds to the reference sample.

It is important to mention that the EPR line could be affected by the "size effect" of the sample that restricts the range of validity of Eq. (1). This effect is observed when the magnetic losses of the sample are important; as a consequence, the quality factor ($Q$) of

the EPR cavity changes at the resonance. [21, 22] In this situation the $I_{EPR}$ is given approximately by:

$$I_{EPR} = \eta \chi_{DC} \omega Q_L \sqrt{1+b} \quad (3)$$

where $\eta$ is the filling factor, $\omega$ is the microwave frequency, $Q_L$ is the loaded $Q$ of the microwave cavity and $b = (4\pi/3) \eta \chi_{DC} \omega Q_L / \gamma \Delta H$, where $\gamma = 2\pi g \mu_B / h$, $\mu_B$ is the Bohr magneton and $h$ is the Plank's constant. The size effect can be made negligible by decreasing the sample mass, in this case $b << 1$ and the intensity is linear with the sample mass.

**2.2 Samples preparation**

The pellets were fabricated by homogenizing stoichiometric proportions of $UO_2$ and $Gd_2O_3$, with a nominal concentration of 7.5% $Gd_2O_3$ over the $UO_2$ total mass. The pellets were pressed in composition batches, where "green" pellets were obtained with a pressure of 300 MPa at room temperature. The pressing process was done in a floating table press, having green densities in the range of 51-53 % of the theoretical density.
The sintering cycle was conducted under a 99.999 % $H_2$ atmosphere, at a flow of 200 ml/min, in a molybdenum furnace, raising the temperature up to 1650 °C in 2 hours and to 1750 °C in 8 hours, respectively. In order to determine the Gd content profile, a cross-section of a pellet of 7 mm side length was cut, and then it was divided into twelve sections of ~2 mm thick, as schematized in Fig. 1. This allows to perform a transversal mapping of $Gd_2O_3$ concentration. In order to compare the EPR properties of the parent compound, $UO_2$ pellets were also fabricated under the same conditions.

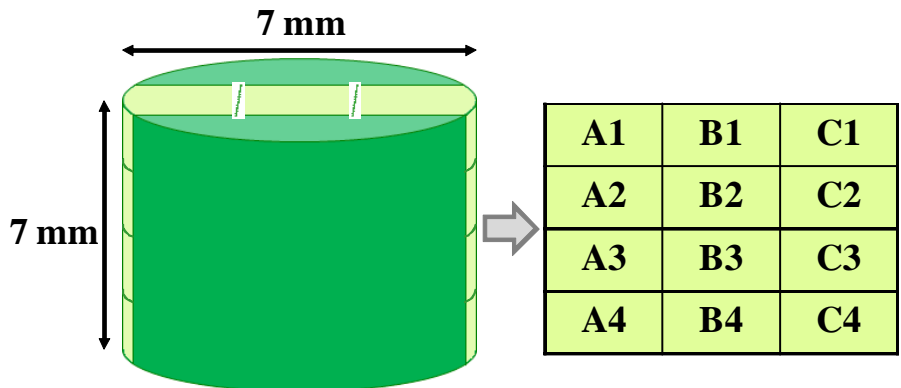

Figure 1: Scheme of the analyzed cross-section of the $UO_2$:$Gd_2O_3$ pellet.

As a reference sample commercial $Gd_2O_3$ (99.99% purity) was employed. In the present case, the gadolinium oxide presents several advantages to be used as reference because it is paramagnetic down to 9 K, presents an intense EPR absorption and makes the comparison direct, i.e., in Eq (2) $g_A=g_O$ and $S_A=S_O$. The $Gd_2O_3$ pellet was prepared by sintering the powder at 1200ºC for 24 hours. The EPR spectra were recorded in a Bruker ESP300 spectrometer operating at X-Band ($\nu = \omega/2\pi \approx 9.5$ GHz). We added a simple accessory to the standard resonant cavity, which allows to control the vertical positioning of the sample holder. All measurements were performed at room temperature and all spectra were normalized by the total sample mass.

## 3. Results and discussion

Figure 2 shows a $Gd_2O_3$ representative spectrum derivative, where a single broad resonance line centered at $H_r \sim 3.33$ kOe is observed. From the resonance field the gyromagnetic factor $g \approx 2.06$ was calculated. The lineshape is mostly determined by the large distribution of Gd-Gd dipolar fields. [23, 24] It can be observed that the line is asymmetrical, with a linewidth which is of the same order of magnitude as the resonance field. This lineshape is caused by the superposition of the spectral lines resonating both at positive and negative fields $\pm H_r$. [25] Taking into account the previous statement, in order to quantify the EPR parameters, the spectral signal was fitted with a derivative of a single Lorentzian line, $L'(H)$, considering also the component centered at negative fields:

$$L'(H) = -\frac{I_{EPR} \Delta H (H - H_r)}{\left[4(H - H_r)^2 + 3\Delta H^2\right]^2} - \frac{I_{EPR} \Delta H (H + H_r)}{\left[4(H + H_r)^2 + 3\Delta H^2\right]^2} \quad (4)$$

In Figure 2 a representative spectrum with the corresponding fitting curve (dotted line) are shown. As described by Eq. (3), the EPR intensity of large samples could be modified by size effects. To avoid this problem the spectrum of $Gd_2O_3$ samples was measured as a function of the mass and the range where $I_{EPR}$ is linear was determined. Figure 3 presents the evolution of $I_{EPR}$ as a function of $Gd_2O_3$ mass. It can be observed that the data exhibit a linear behavior up to a mass m~30 mg. Therefore a sample with m=(4.5±0.2) mg was chosen as the reference sample.

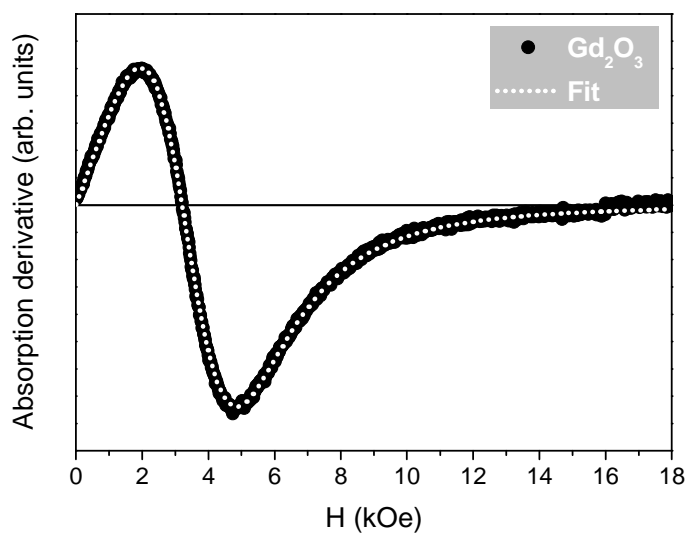

Figure 2: EPR spectrum of the $Gd_2O_3$ reference sample with a mass m=(4.5±0.2) mg (solid line) and fitting curve (dotted line).

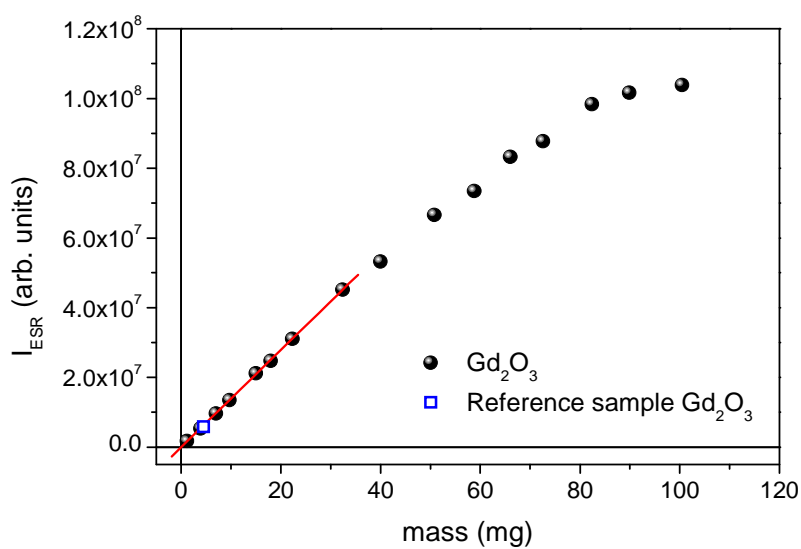

Figure 3: Calibration curve of the EPR intensity as a function of $Gd_2O_3$ mass. The straight line indicates the range where the $I_{EPR}$ signal exhibits a linear dependence with the sample mass.

Once the reference sample was obtained, we proceeded with the analysis of the cross-section of the fuel pellet. A typical spectrum, corresponding to the A4 section, is presented in Fig. 4. This figure also includes the spectrum of a pure $UO_2$ ceramic sample. Notice that the $UO_2$ line presents negligible EPR intensity compared to the $UO_2$-$Gd_2O_3$ pellets. This fact allowed us to identify the $UO_2$-$Gd_2O_3$ signal as coming almost exclusively from the Gd contribution. In this case it was necessary to fit the $UO_2$-$Gd_2O_3$ spectral line with two Lorentzian lines (considering also their respective components centered at negative fields) with almost the same *g*-value but different linewidths. These lines could be originated by $Gd^{3+}$ ions located at different crystalline sites. In particular, the broader line could be signaling the presence of significant dipolar interactions and $Gd_2O_3$ "clustering", as it was previously observed for other $Gd^{3+}$ systems [26]. The $I_{EPR}$ parameter was obtained from the total fitting curve (see dotted curve in Fig. 4).

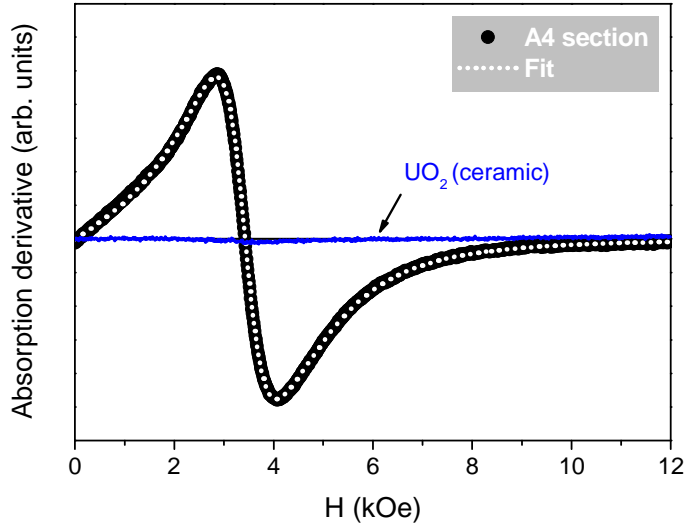

Figure 4: ESR spectrum corresponding to the A4 section of the $UO_2$-$Gd_2O_3$ pellet (solid line). The dotted line is the fitting curve, considering two Lorentzian lines and their respective components centered at negative fields, and the arrow signals the ESR spectrum corresponding to a $UO_2$ ceramic sample.

Finally, from Eq. 2, the Gd concentration in each section of the pellet was calculated by comparing their EPR intensities with the $Gd_2O_3$ reference sample. In Table 1 the percentual concentration of Gd ions per gram of compound of each section in the pellet is presented and this profile is schematized in Fig. 5. On the other hand, if it is assumed that all Gd ions in the sample are in the $Gd_2O_3$ phase, the percentage of $Gd_2O_3$ in each section can be quantified. This result is summarized in Table 2. Although the average $Gd_2O_3$ percentual concentration obtained is equal to the nominal value of gadolinium oxide employed in the fabrication process (7.5%), it can be observed that there is a gradient of concentrations: the $Gd_2O_3$ is more concentrated in the center of the pellet (with a maximum measured value of 9.3%) and diminishes towards the edges (where a minimum value of 5.8% was measured for the C4 section).

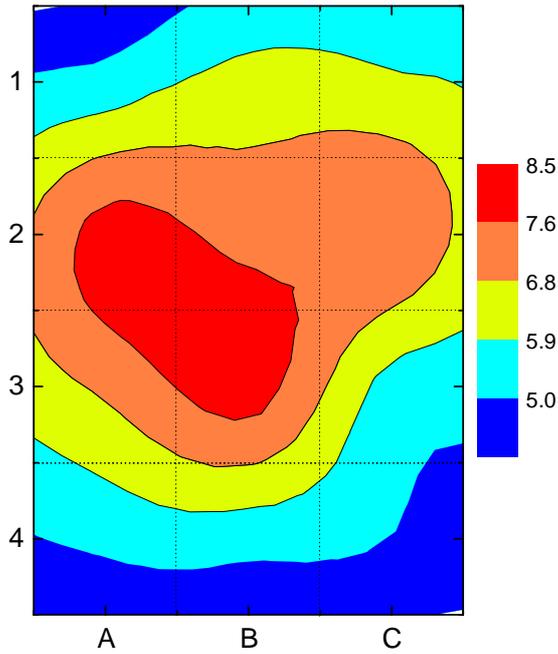
Figure 5: schematic illustration of the Gd concentration profile, based on a linear interpolation of the experimental results presented in Table 1. The chromatic scale indicates the concentration gradient.

|   | A | B | C |
|---|---|---|---|
| 1 | 5.3 ± 0.3 | 6.2 ± 0.3 | 6.0 ± 0.3 |
| 2 | 8.0 ± 0.4 | 7.3 ± 0.4 | 7.6 ± 0.4 |
| 3 | 6.9 ± 0.4 | 8.1 ± 0.5 | 5.5 ± 0.3 |
| 4 | 5.2 ± 0.3 | 5.3 ± 0.3 | 5.0 ± 0.3 |

Table 1: percentual concentration of Gd ions per gram of compound at each analyzed section of the pellet.

|   | A | B | C |
|---|---|---|---|
| 1 | 6.2 ± 0.3 | 7.2 ± 0.4 | 7.0 ± 0.4 |
| 2 | 9.3 ± 0.5 | 8.5 ± 0.5 | 8.8 ± 0.5 |
| 3 | 8.0 ± 0.5 | 9.3 ± 0.6 | 6.3 ± 0.3 |
| 4 | 5.9 ± 0.3 | 7.3 ± 0.3 | 5.8 ± 0.3 |

Table 2: percentage of $Gd_2O_3$ mass concentration with respect to the total mass at each analyzed section of the pellet.

## 4. Conclusions

In this work we have shown that EPR spectroscopy could be employed to perform quality control of $UO_2$-$Gd_2O_3$ fuel pellets. From the comparison with a reference sample, the Gd concentration in a cross-section of a $UO_2$-$Gd_2O_3$ fuel pellet was quantified. We presented a simple measurement protocol that allows to determine the relative concentration of Gd in $UO_2$ with an error smaller than 5%. In the studied sample we observed that the distribution of Gd atoms is not uniform. The magnetic ions are more concentrated at the center of the pellet and this concentration diminishes towards the edges. Assuming the Gd ions are in the $Gd_2O_3$ phase, the average concentration coincides with the nominal concentration of $Gd_2O_3$ employed in the pellet fabrication. However, the local concentration varies from 5.8% to 9.3%.

Finally we would like to emphasize that the ESR spectroscopy, with a proper measurement protocol, can be applied as a calibration standard method. In particular, it would allow to validate other routine quality control techniques of easier implementation in the fabrication process.

**References**

[1] S.J. Zinkle, G.S. Was, Acta Materialia 61 (2013) 735.

[2] M. Durazzo, A.M. Saliba-Silva, E.F. Urano de Carvalho, H.G. Riella, J. Nucl. Mater. 433 (2013) 334.

[3] K. W. Song, K. S. Kim, J. H. Yang, K. W. Kang, Y. H. Jung, J. Nucl. Mater. 288 (2001) 92.

[4] A.G. Leyva, D. Vega, V. Trimarco, D. Marchi, J. Nucl. Mater. 303 (2002) 29.

[5] L. Pagano Jr., G.P. Valença, S.L. Silva, A.E.L. Cláudio, F.F. Ivashita, R. Barco, S.N. de Medeiros, A. Paesano Jr., J. Nucl. Mater. 378 (2008) 25.

[6] K. S. Kim, J. H. Yang, K. W. Kang, K. W. Song, G. M. Kim, J. Nucl. Mater. 325 (2004) 129–133

[7] K.W. Kang, J.H. Yang, J.H. Kim, Y.W. Rhee, D.J. Kim, K.S. Kim, K.W. Song, Thermochimica Acta 455 (2007) 134.

[8] K. Iwasaki, T. Matsui, K. Yanai, R. Yuda, Y. Arita, T. Nagasaki, N. Yokoyama, I. Tokura, K. Une, K. Harada, J. Nucl. Sci. Technol. 46 (2009) 673.

[9] F. Zaldivar Escola, O.E. Martínez, N. Mingolo, R. Kemp, J. Nucl. Mater. 435 (2013) 17–24.

[10] L. Halldahl and S. Eriksson, J. Nucl. Mater. 153 (I 988) 66-70

[11] S. K. Misra (Ed.), Multifrequency Electron Paramagnetic Resonance: Theory and Applications (Wiley-VCH, 2011)

[12] J. A. Weil and J. R. Bolton, Electron Paramagnetic Resonance, Elementary Theory and Practical Applications (John Wiley & Sons, Inc, 2007).

[13] T.-T. Chang, D. Foster, A.H. Kahn, Journal of Research of the National Bureau of Standards 83 (1978) 133

[14] A. Abraham and B. Bleaney, Electron Paramagnetic Resonance of Transition Ions (Clarendon Press, Oxford, 1970)

[15] ISO / ASTM51607 - 04 Standard Practice for Use of the Alanine-EPR Dosimetry System (2004).

[16] C. Fainstein, E. Winkler, M. Saravi, Appl. Rad. and Isotopes 52 (2000) 1195.

[17] M. Anton Phys. Med. Biol. 51 (2006) 5419–5440

[18] A. Lund, M. Shiotani and S. Shimada, Principles and Applications of ESR Spectroscopy (Springer Science+Business Media B.V., 2011).

[19] C. Miyake, M. Kanamuro and S. Imoto, J. Nuc. Mat. 137 (1986) 256

[20] C. P. Poole, Electron Spin Resonance: A Comprehensive Treatise on Experimental Techniques (Courier Dover Publications, 1996),

[21] M. S. Seehra, Rev. Sce. Instrum. 39 (1968) 1044.

[22] M. T. Causa, M. Tovar, A. Caneiro, F. Prado, G. Ibañez, C. A. Ramos, A. Butera, B. Alascio, X. Obradors, S. Piñol, F. Rivadulla, C. Vázquez-Vázquez, A. López-Quintela, J. Rivas, Y. Tokura, and S. B. Oseroff, Phys. Rev. B 58 (1998) 3233.

[23] R.S. de Biasi and M. L. N. Grillo, J. Phys. Chem. Solids 65 (2004) 1207.

[24] A. Fainstein, E. Winkler, A. Butera and J. Tallon, Phys. Rev. B 60 (1999) R12597

[25] This is a well-known effect in usually encountered in magnetically concentrated materials and polycrystalline samples. See, for example, M. T. Causa and M. C. G. Passeggi, Phys. Rev. B 32 (1985) 3229.

[26] A. I. Smirnov and S. Sen, J. Chem. Phys. 115 (2001) 7650.